\documentclass[referee]{raa}

\usepackage{graphicx,times,ulem,titletoc}             
\usepackage{lscape}
\usepackage{url}
\usepackage{caption}
\begin{document}

\title{The First Data Release of LAMOST Low Resolution Single Epoch Spectra }

\volnopage{Vol.0 (200x) No.0, 000--000}      
\setcounter{page}{1}          

\author{Zhong-Rui Bai      \inst{1,2}
   \and Hao-Tong Zhang      \inst{1,2}
   \and Hai-Long Yuan      \inst{1,2}
      \and Dong-Wei Fan \inst{2,3}
      \and Bo-Liang He \inst{2,3}
   \and Ya-Juan Lei      \inst{1,2}
   \and Yi-Qiao Dong      \inst{1,2}
   \and Si-Cheng Yu      \inst{1,2}
   \and Yong-Heng Zhao \inst{1,2}
\and Yong Zhang \inst{4}
\and Yong-Hui Hou \inst{4}
   \and Yao-Quan Chu \inst{5}
}
 
\institute{
$^{1}$CAS Key Laboratory of Optical Astronomy, National Astronomical Observatories, Beijing 100101, China; {\it htzhang@bao.ac.cn}\\
$^{2}$National Astronomical Observatories, Chinese Academy of Sciences, Beijing 100101, China\\
$^{3}$National Astronomical Data Center, Beijing 100101, China\\
$^{4}$Nanjing Institute of Astronomical Optics \& Technology, National Astronomical Observatories, Chinese Academy of Sciences, Nanjing 210042, China\\
$^{5}$University of Science and Technology of China, Hefei 230026, China
}

\date{Received 2021 March 29; accepted 2021 May 17}

\abstract{LAMOST Data Release 5, covering $\sim$17000 $deg^2$ from $-10^{\circ}$ to $80^{\circ}$ in declination, 
contains 9 million co-added low-resolution spectra of celestial objects, each spectrum combined from repeat exposure of  two to tens of times  during Oct 2011 to Jun 2017. 
In this paper, we present the spectra of individual exposures for all the objects in LAMOST Data Release 5. 
For each spectrum, the equivalent width of 60 lines from 11  different elements are calculated with a new method combining  the actual line core and fitted line wings.  
For stars earlier than F type,  the Balmer lines are fitted with both emission and absorption profiles once two components are detected.   
Radial velocity of each individual exposure is measured by minimizing ${\chi}^2$  between the spectrum and  its best template. 
The database for equivalent widths of spectral lines and radial velocities of individual spectra are available online. 
Radial velocity uncertainties with different stellar type and signal-to-noise ratio are quantified by comparing different exposure of the same objects. 
We notice that the radial velocity uncertainty  depends on the time lag between observations. 
For stars observed in the same day and with signal-to-noise ratio higher than 20, the radial velocity uncertainty  is  below 5 km s$^{-1}$, and increase to 10 km s$^{-1}$ for stars observed in different nights.
}

\authorrunning{Z.-R. Bai et al. } 
\titlerunning{LAMOST single epoch DR}  

\maketitle

\section {Introduction}

With the increasing number of time domain astronomical surveys in the past two decades, 
the samples of various types of variable stars are increasing rapidly. 
Photometric surveys, such as Kepler (\cite{kepler2010}), OGLE (\cite{OGLE2015}), Catalina (\cite{catalina2014}), ASAS-SN (\cite{asas2014}) and TESS (\cite{TESS2016}), 
have obtained millions of high-precision stellar light curves, 
providing the opportunity to discover  a large number of variables such as eclipsing binaries. 
The corresponding spectra could help to constrain the physics of these variables.
Furthermore, their  variation in radial velocity (RV) or  spectral features  provide more information to better determine their property and further explore their nature. 
RV, derived from the spectra, provides information about pulsation, orbital motion of binaries, as well as the expansion of eruptive variables such as supernovas, etc. 
RV analysis is a necessary and reliable tool to detect and confirm these types of objects.
In addition to RV, the variation of  spectral features can also provide  clues of finding some special objects, such as  accretion disk in compact objects and shock waves in pulsation stars, etc.  These time-domain spectral information could also provide an independent way to study   variables even without time-domain photometric results. 

Most spectra of the on-going multi-fiber spectroscopic surveys are observed multiple times in order to improve the signal-to-noise ratio (SNR)  and remove cosmic rays in the combined spectra.
The individual exposures  naturally provide time-domain information about the source. 
Previous work based on single epoch spectra like RAVE (\cite{matij2011}), SDSS (\cite{Bickerton2012}) and APOGEE (\cite{APOGEE2020}) mostly focus on binaries. 

Among all the spectroscopic surveys, the LAMOST survey (\cite{cui2012}, \cite{zhao2012}) has released by far the largest number of spectra in the world.
In LAMOST Data Release 5 (DR5), the data release of the first stage of LAMOST survey, the General Catalog (hereafter DR5GC) published 9 million spectra, 8 million of which are stellar spectra.
Works have been done using the published RV of individual objects.
For example, \cite{qian2019} found more than 250000 spectroscopic binary or variable star candidates with RV difference larger than 10 km s$^{-1}$. 
\cite{gao2017} estimated the binary fraction based on the RV  of all the repeated LAMOST observations. 
\cite{liu2019} and \cite{gu2019} made use of the RV variation to find  stellar black hole candidates.

However, these results are all based on the RV from the co-added spectrum of several single exposures in the same observation, so that the time scale is usually longer than one day.
Results with a shorter time scale in tens of minutes can be obtained by spectral variation between single exposures. 
In order to make full use of LAMOST data, we release the corresponding single epoch data of all sources in LAMOST DR5GC. 
The data consist of three parts: spectra data, RV data and spectral line equivalent width (EW) data.
This collection is a treasure for studying binaries and other variable stars in a shorter time scale.
The single epoch spectra of galaxies and QSOs are also provided in this work, but no RV or EW measurement is attempted.   
 
In Section 2, we introduce LAMOST DR5 and the data reduction.
The measurements of EW and RV are described in Section 3 and Section 4, respectively.
The data description is listed in Section 5. Section 6 gives a summary.

\section {LAMOST survey}
\subsection{LAMOST DR5}
\begin{figure}[bthp]
\begin{minipage}{\textwidth}
\centering
\includegraphics[width=0.95\textwidth]{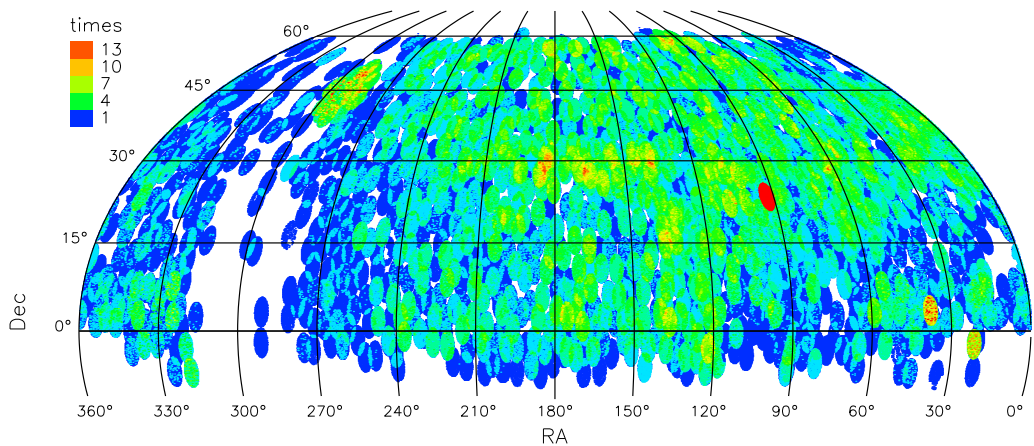}
\caption{The footprint of LAMOST DR5. The number of repeated observations for each target are indicated by different colors, as shown in the color bar. \protect\footnotemark}
\label{footprint}
\end{minipage}
\end{figure}

LAMOST DR5 (\url{http://dr5.lamost.org/}) is the latest data release before the second stage of the LAMOST survey. The data, including the pilot survey (October 2011 to June 2012) and the first stage of sky survey (September 2012 to June 2017), 
covers $\sim$17000 deg$^2$ from $-10^{\circ}$ to $80^{\circ}$ in declination, as shown  in Figure \ref{footprint}.

With its 4000 fibers, LAMOST is able to observe up to 4000 targets simultaneously. 
A LAMOST plate is a configuration of 4000 fibers consisting of  target fibers, sky fibers and empty fibers. 
According to the  brightness of the targets, LAMOST plates are grouped into V (very bright, $r\leq14$), B (bright, $14<r<16.8$), M (medium, $16.8\leq{r}\leq{17.8}$) and F (faint, $r>17.8$). 
The typical single exposure durations for V, B, M, F plates are 600, 900, 1200 and 1800 seconds, respectively.
 Usually each plate is exposed consecutively 2-3 times in the same night to remove cosmic rays and other artifacts. 
Throughout this paper, these consecutive exposures of the  same plate in the same night are defined as one observation. 
Thus, the same target in two plates is considered to be observed in two observations rather than one even if the two plates were observed in the same night.  

\begin{figure}[bthp]
\centering
\includegraphics[width=0.5\textwidth]{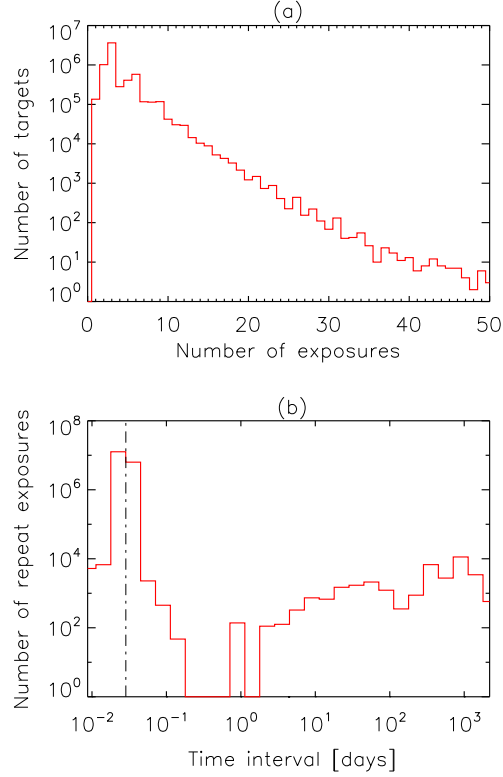}
\caption{Upper: The histogram of the target number corresponding to the number of repeated exposures in LAMOST DR5. Lower: The histogram of the time interval between exposures. 
The peak indicated by the dash-dotted line in the lower panel is half an hour, corresponding to the time interval between LAMOST adjacent exposures of the same plate.}
\label{obssta}
\end{figure}

Each spectrum released in DR5 is the combination of individual epoch spectra in one observation.  
Spectra of individual exposures and the corresponding RV and EW measurements were not released.
Differing from the ongoing second stage survey, which is a hybrid of low and medium resolution for both regular and time domain survey, 
spectra in DR5 were observed in low resolution mode of R$\sim$1800 and mostly not designed for time domain survey. 
Most of the targets were repeated occasionally, usually  less than 10 times, as can be appreciated from Figure \ref{footprint}.  
Figure \ref{obssta} shows distribution of the number of individual exposures  and the time intervals between exposures for repeated observations of a single target. 
Most of the targets had no more than three exposures in less than 2 hours, i.e., only one observation. 
127 thousands targets have more than 10 exposures. 
Most of the exposures are repeated within 4 hours, and for about 10000 targets, the time lags are longer than 3 years.

\footnotetext{ Note in the figure, the plate (RA=$92^{\circ}.9844$, Dec=$23^{\circ}.2071$) with highest observation numbers is a special designed time domain plate in the Kepler K2 campaign 0 field. 
Since the plate was observed from 2016 to 2019, which is beyond the scope of current data release, for the completeness of the time series data, the plate will be released separately in another paper, rather than in the current data release.}

\begin{figure}[bthp]
\centering
\includegraphics[width=0.45\textwidth]{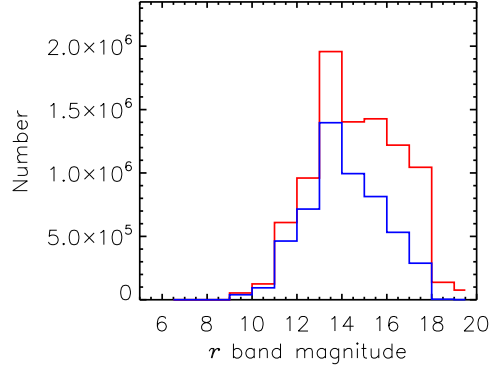}
\caption{The distribution of {\it r} magnitude in LAMOST DR5. Red and blue are for  the general catalog and parameter catalog respectively. }
\label{magdis}
\end{figure}

\begin{figure}[bthp]
\centering
\includegraphics[width=0.45\textwidth]{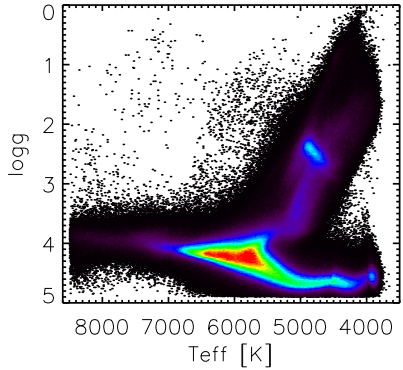}
\caption{$\mathrm{log}\textit{ g}$ versus $T_{\mathrm{eff}}$ of 5 millions stars in DR5PC. Color coded parts are relative number density.
The hotter the color, the higher the density.}
\label{t2logg}
\end{figure}

There are 9 million flux and wavelength calibrated, sky-subtracted spectra in DR5, including 8 million stars. 
Stars were observed  in all spectral type and the range of magnitude in $r$ band is $9 \sim 20$ (Figure \ref{magdis}).  
For 5.3 million of them, mostly of which are A, F, G, K type stars within the temperature range of 3700 $\sim$ 8500 K, 
the basic stellar parameters such as effective temperature ($T_{\mathrm{eff}}$), surface gravity ($\mathrm{log}\textit{ g}$), metallicities ($[\mathrm{Fe/H}]$) and RVs
are provided in the DR5 Parameter Catalog (hereafter DR5PC), while for the rest stars in DR5GC, the stellar parameters are not available.
$T_{\mathrm{eff}}$, $\mathrm{log}\textit{ g}$ and $[\mathrm{Fe/H}]$ cover the range of 3700 $\sim$ 8500 K, 0 $\sim$ 5 dex and -2 $\sim$ 1 dex, respectively.  
Comparing with APOGEE data, the error of $T_{\mathrm{eff}}$, $\mathrm{log}\textit{ g}$ and $[\mathrm{Fe/H}]$ for the stars with SNR greater than 100 are 70 K, 0.14 dex and 0.06 dex, respectively, deteriorating to 107 K, 0.23 dex and 0.1 dex for stars with SNR in the range of 30 to 50 \footnotemark.   
Figure \ref{t2logg} shows the distribution of stars in  the HR  diagram ($T_{\mathrm{eff}}$ vs $\mathrm{log}\textit{ g}$) of DR5PC.
\footnotetext{http://dr5.lamost.org/v3/doc/release-note-v3}

\subsection{Data Reduction}
\begin{figure}[bthp]
\centering
\includegraphics[width=0.5\textwidth]{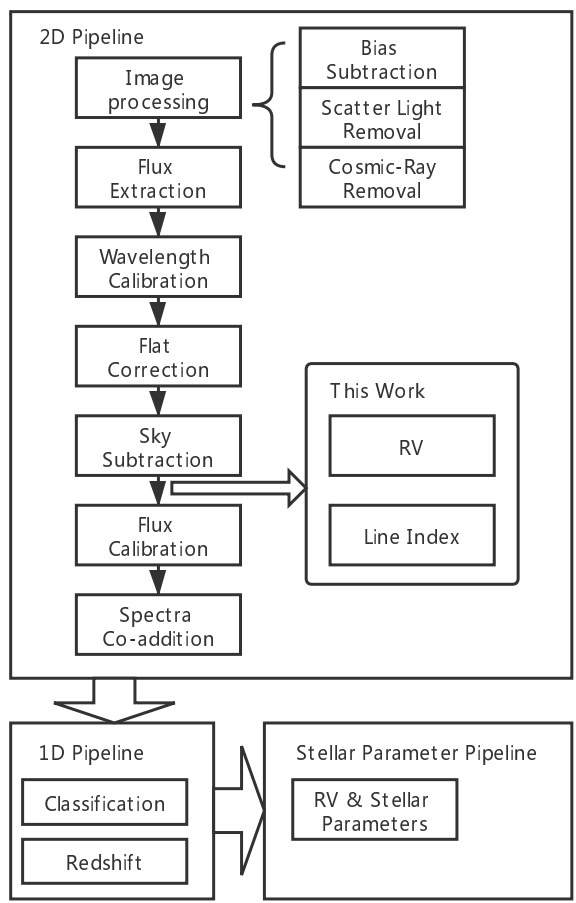}
\caption{Flow chart of LAMOST data reduction pipeline. The process related to the current work is marked by a mini panel.}
\label{pipeline}
\end{figure}

Lights from 4000 fibers are fed into 16 spectrographs, each holding 250 fibers. In the low resolution mode, the light beam is split into blue (3700-5900\AA) and red (5700-9000\AA)  by a  dichroic mirror in each spectrograph. 
Each segregated beam is then dispersed by a volume phased holographic grating and finally focused on a 4K$\times$4K CCD through a Schmidt camera respectively. 
Since the light path  thus the instrument effect is different for the targets on different CCD, each CCD image is processed independently in the data reduction pipeline.  
The raw data are first processed by the LAMOST two-dimensional data reduction pipeline (2D pipeline), which is a part of LAMOST data pipeline. 
The basic data reduction steps and the relationship between different parts of the pipeline  are laid out in  the flow chart of Figure \ref{pipeline}. 

Each  CCD image is first bias subtracted,  then cosmic rays  are rejected by  a two-dimensional profile fitting method (\cite{bai2017a}). 
For each row in the spatial direction, the scatter light is sampled in  the pixels of dead fibers and those between the fiber trace   then interpolated to all the pixels. The final scatter light is filtered by a 5 by 5
median boxcar and subtracted from the 2D image. This step is applied to both the flat field and object images. 
The fiber traces are first derived from the flat field image taken at twilight in the zenith direction. 
Then for each individual exposure image,  the fiber trace is adjusted in the spatial direction  to better match the profile center in the object image.  
Spectrum for each fiber is then extracted by integrating a weighted aperture of 15 pixels  along the corrected fiber trace.
Arc lamps (Hg/Cd for blue and Ne/Ar for red) taken at the beginning, middle and the end of the observation night are used to calibrate the wavelength. 
Wavelength solutions are interpolated to derive the wavelength of the spectra taken between the arc lamps. 
The wavelength is further tweaked by the sky emission lines in both red and blue spectra. 
The overall wavelength calibration accuracy is less than 5 km s$^{-1}$ (\cite{bai2017b}). 
The final output wavelength is corrected to  the heliocentric frame in vacuum.
The fiber-to-fiber relative throughput are corrected by the flat field  and further adjusted by sky emission lines. 
The sky background spectrum built from  sky fibers is subtracted from object spectra after careful wavelength and throughput correction.    
The detail of wavelength calibration, throughput correction  and sky subtraction algorithm could be found in \cite{bai2017b}. 

In DR5, the flux calibration curve is obtained through the selected F-type flux standard stars and their corresponding templates. The blue and red sky-subtracted spectra are divided by their respective curves, and then stitched together.
The results of the 2D pipeline, i.e., the co-added spectra, are then analyzed by the LAMOST one-dimensional data pipeline (1D pipeline) and the LAMOST Stellar Parameter pipeline (LASP, \cite{wu2011}) to derive their spectral type and basic parameters.
The single epoch spectra released in this work are the intermediate produce of the 2D pipeline. 
The spectra in  blue and red are released separately and not flux calibrated.  
We attached the flux calibration curve in the data product (see section \ref{sec:product}), so the flux could be relatively corrected   if necessary.
The measurements of RV and EW are discussed in the following sections. 

\section {Spectral line measurement}
The variation of a spectral line can reveal physical phenomena, such as the stellar activity, the variation of the gas around the star, or the motion of the star itself.
EW, which is less affected by spectral resolution, can be used to compare  lines of different observation time and different wavelength.
In a certain range, EW of some spectral lines can be used as indicators of stellar parameters.
For example, the EW of H$\alpha$ varies obviously with the effective temperature widely across the star types.
In this work, we select a series of frequently-used lines and calculate their EWs and RVs in single epoch data. 
\subsection {List of lines}
As many as 60 elemental lines are measured, 
whose wavelengths and sources are listed in Table \ref{linelist}.
Some blended lines are not measured because they cannot be independently separated at the low resolution mode ($\Delta\lambda$=2.8\AA\ @ 5000\AA\  and $\Delta\lambda$=4.4\AA\ @ 8000\AA), 
such as Mg lines ($\lambda5167, 5172, 5183$), Na doublets ($\lambda5890, 5896$ and $\lambda8183, 8194$).
The blended lines will be measured with an alternative method in the future work.
To be mentioned, He $\lambda3888$/H $\lambda3889$ and Fe $\lambda4920$/He $\lambda4921$ have the close wavelength values that we treat them as single lines.
We do not judge which elements they come from or decompose them.

\begin{table}[bthp]
\begin{center}
\centering \caption{List of the lines measured in this work.} \label{linelist}
\begin{tabular}{cll|cll|cll}
\hline\hline
Line & Central (\AA)& Description  & Line & Central (\AA)& Description  & Line & Central (\AA)& Description \\
\hline
3835 $^s$ & 3835.384    &   H$\epsilon$    & 4388 $^s$ & 4387.929    &   He I           & 6583 $^n$ & 6583.45     &   N II           \\
3888 $^s$ & 3888.648    &   He I           & 4471 $^s$ & 4471.480    &   He I           & 6678 $^s$ & 6678.15     &   He I           \\
3889 $^s$ & 3889.064    &   H$\epsilon$    & 4481 $^s$ & 4481.33     &   Mg II          & 6716 $^n$ & 6716.44     &   S II           \\
3933 $^s$ & 3933.663    &   Ca K           & 4541 $^s$ & 4541.6      &   He II          & 6730 $^n$ & 6730.816    &   S II           \\
3968 $^s$ & 3968.469    &   Ca H           & 4554 $^s$ & 4554.033    &   Ba II          & 6841 $^s$ & 6841.339    &   Fe I           \\
3970 $^s$ & 3970.075    &   H$\epsilon$    & 4686 $^s$ & 4685.7      &   He II          & 6978 $^s$ & 6978.851    &   Fe I           \\
3995 $^s$ & 3995.000    &   N II           & 4713 $^s$ & 4713.146    &   He I           & 7065 $^s$ & 7065.19     &   He I           \\
4009 $^s$ & 4009.256    &   He I           & 4861 $^b$ & 4861.35     &   H$\beta$       & 7389 $^s$ & 7389.398    &   Fe I           \\
4026 $^s$ & 4026.191    &   He I           & 4920 $^s$ & 4920.502    &   Fe I           & 7748 $^s$ & 7748.269    &   Fe I           \\
4046 $^s$ & 4045.812    &   Fe I           & 4921 $^s$ & 4921.931    &   He I           & 8435 $^s$ & 8435.650    &   Ti I           \\
4058 $^s$ & 4057.760    &   N IV           & 4957 $^s$ & 4957.597    &   Fe I           & 8468 $^s$ & 8468.407    &   Fe I           \\
4063 $^s$ & 4063.594    &   Fe I           & 4959 $^n$ & 4958.711    &   O III          & 8498 $^s$ & 8498.02     &   Ca II tri.     \\
4071 $^s$ & 4071.738    &   He I           & 5007 $^n$ & 5006.843    &   O III          & 8515 $^s$ & 8515.108    &   Fe I           \\
4077 $^s$ & 4077.140    &   Sr II          & 5015 $^s$ & 5015.678    &   He I           & 8542 $^s$ & 8542.09     &   Ca II tri.     \\
4102 $^s$ & 4101.734    &   H$\delta$      & 5047 $^s$ & 5047.74     &   He I           & 8598 $^s$ & 8598.39     &   P14            \\
4121 $^s$ & 4120.816    &   He I           & 5227 $^s$ & 5227.150    &   Fe I           & 8662 $^s$ & 8662.14     &   Ca II tri.     \\
4200 $^s$ & 4199.9      &   He II          & 5411 $^s$ & 5411.52     &   He II          & 8665 $^s$ & 8665.02     &   P13            \\
4226 $^s$ & 4226.73     &   Ca I           & 6494 $^s$ & 6494.981    &   Fe I           & 8688 $^s$ & 8688.625    &   Fe I           \\
4340 $^s$ & 4340.472    &   H$\gamma$      & 6548 $^n$ & 6548.05     &   N II           & 8750 $^s$ & 8750.46     &   P12            \\
4383 $^s$ & 4383.545    &   Fe I           & 6563 $^b$ & 6562.79     &   H$\alpha$      & 8863 $^s$ & 8862.89     &   P11            \\
\hline
\end{tabular}
\end{center}
$^s$ Stellar spectral lines.\\
$^b$ Stellar spectral lines or nebular emission lines.\\
$^n$ Nebular emission lines.
\end{table}

\subsection {Measurement}\label{secnmlines}
Given a spectrum from LAMOST single epoch data, it is first normalized to remove the continuum.
A certain range around each line is fitted by a symmetric function to measure the line center and calculate the full width at half maxima (FWHM).
Then the spectrum within a certain width is added up to obtain EW.
The details are as follows.

\textbf { Normalization. }
Blue and red branches are normalized separately.
For each branch, the whole spectrum is first median-filtered by a 9-pixel window to remove the outliers.
Then a 15-order polynomial, iteratively fitted with both absorption and emission lines rejected, 
is taken as the continuum and divided.

\textbf { Fitting range.}
We use a symmetry function to fit each line.
Before that, we need to determine the range of fitting.
Since the width of the lines varies between stars by temperature, gravity and rotation, etc, there is no  universally 
applicable range to measure all the lines in different types of stars. To avoid the complexity of measuring any single line
with a set of fixed window size,  we develop a  method  to  measure the EW of lines with an adaptive window size.   
To determine the proper fitting range, the normalized spectrum is temporarily smoothed by a median-filter with the width of 5 pixels. 
For a line with the wavelength of $\lambda$,
the range from $\lambda-1.5\lambda_r$\ to\ $\lambda+1.5\lambda_r$ is initially selected, where $\lambda_r$ is the resolution of the spectrum.
For LAMOST, the resolution is $R\sim$1800, so we adopt 3.0\AA\ and 4.5\AA\ for blue and red branches, respectively.
The boundaries of the initial range continue to grow until the flux begins to significantly cease to increase for an absorption line or decrease for an emission line, and the final range is determined.

\textbf { Fitting. }
The segment of the normalized spectrum within the fitting range is then fitted with 
a S\'ersic profile (\cite{ser68}) added by 1, as described in Eq. \ref{seric}, 
where $a$, $\lambda_0$ is the intensity and center of the line, $\sigma$ and $\delta$ are two parameters.
The Pearson correlation coefficient (CC) between the spectrum
and the fitted profile is calculated when the fit is successful. 
Generally, CC greater than 0.9 indicates that the fit is reliable while lower CC means worse fitting.
 Since the calculation of EW is based on the fitted profile (see below), EW is not calculated  if the fit fails or the Pearson CC is less than 0.7.
\begin{equation}\label{seric}
f(\lambda)=a e^{-\frac{|\lambda-\lambda_0|^\delta}{\delta\sigma^{\delta}}}+1\end{equation}

\textbf { Calculation.}
Once the line is successfully fitted, FWHM is calculated.
EW is defined as the integration of the normalized spectrum subtracted by 1 over the wavelength region $|\lambda-\lambda_0|<2\lambda_{\mathrm{FWHM}}$.
The line wing beyond this region is not  used due to its much less contribution and larger uncertainty. 
The integration is consist of two parts.  
For the  line core  ($|\lambda-\lambda_0|<\lambda_{\mathrm{FWHM}}$),
the integration is performed on the actual spectrum to account for the possible second component (e.g. emission core).
While for the outer part of the line ($\lambda_{\mathrm{FWHM}}\leq|\lambda-\lambda_0|<2\lambda_{\mathrm{FWHM}}$), 
the fitted profile is used instead to calculate the integration to avoid the influence of weaker lines of other elements.
If the line is an emission, the EW is set to be negative.
The error of the EW is calculated based on the error cumulation formula.

\subsection {Comparing with fixed width}

\begin{figure}[bthp]
\centering
\includegraphics[width=0.9\textwidth]{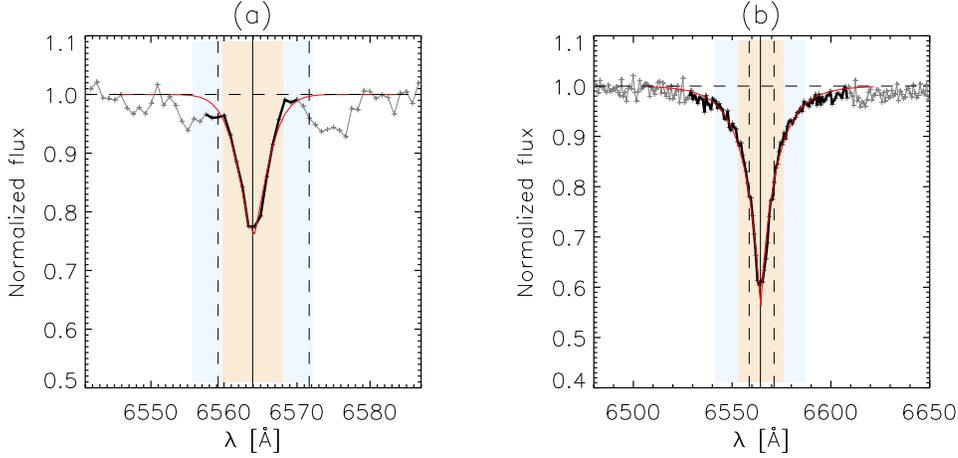}
\caption{Two examples of spectra line fitting. 
(a) and (b) show the fit of H$\alpha$ of a K7 type star and an A5 type star, respectively.
The continuum-normalized spectra are shown in grey and the fitted line profiles are in red.
The dash vertical lines limit the initial boundary and spectra within the final boundary is in black.
The solid vertical lines are the centers. The light yellow and light blue areas denote $|\lambda-\lambda_0|<\lambda_{\mathrm{FWHM}}$ and $\lambda_{\mathrm{FWHM}}<|\lambda-\lambda_0|<2\lambda_{\mathrm{FWHM}}$, respectively.
} \label{linefit}
\end{figure}

Considering the huge difference of line width  of different elements  due to different broadening effect (e.g. thermal, pressure and rotation), it is not possible to use a uniform width to measure  EWs for different elements lines in various stellar types. 
There are two ways to meet this challenge: using an adaptive  width like in this work, or using a series of fixed wavelength width, as used in the SSPP (the SUGUE Stellar Parameter Pipeline, \cite{lee2008}). 
Figure \ref{linefit} gives two examples, for a K7 star and an A5 star, respectively.
For the late type stars, since the Balmer lines are relatively weak, the wings contribute much less than the core to the EW, so the adaptive method could exquisitely choose the width neither too wide to contain nearby lines nor too narrow to miss any core region.
The same is true for the early type stars, where the Balmer lines are much stronger and there are less metal lines in the wing.
The wavelength region to calculate EW ($|\lambda-\lambda_0|<2\lambda_{\mathrm{FWHM}}$) looks narrow for early type stars, leading the  EWs to be underestimated but with a certain fraction to the real EW.  
 We did not try to include more wings for the Balmer lines, since the wings may be affected by both other lines and the uncertainty in the continuum determination especially when the resolution is low.   
 
\begin{figure}[bthp]
\centering
\includegraphics[width=0.7\textwidth]{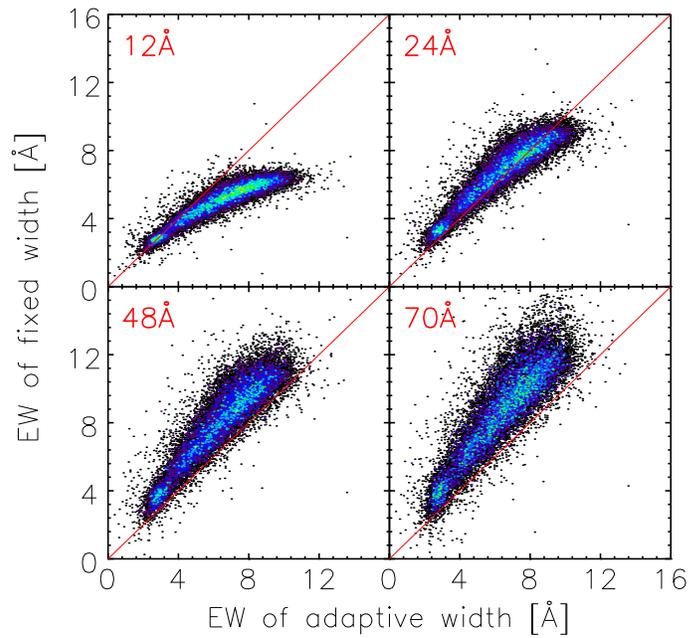}
\caption{The comparison of H$\alpha$ EW for A type stars. 
The y-axis of four panels are EW of fixed width of 12\AA, 24\AA, 48\AA and 70\AA , respectively.
} \label{ewhadif}
\end{figure}

\begin{figure}[bthp]
\centering
\includegraphics[width=0.7\textwidth]{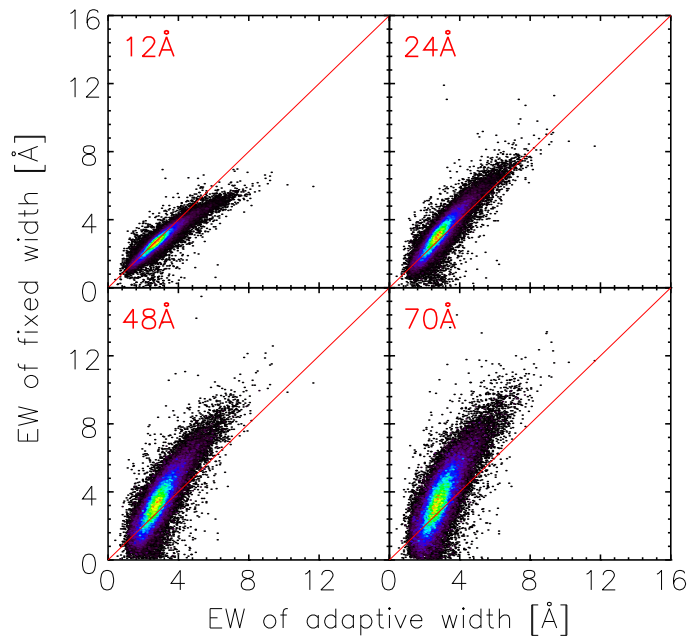}
\caption{The same as Figure \ref{ewhadif} but for F type stars.
} \label{ewhadiff}
\end{figure}

In Figure \ref{ewhadif} and Figure \ref{ewhadiff}, the comparison between EW from adaptive width (hereafter EWaw) and EW of four fixed widths, i.e., 12\AA, 24\AA, 48\AA\ and 70\AA\ (same as the width selected in \cite{lee2008}, hereafter EW12, EW24, EW48 and EW70, respectively), of H$\alpha$ lines from A and F star are present. 
From the figures, we can see that  scatter increases with the increasing of the width, caused by the  pollution of metallic lines and the uncertainty in the continuum determination of the line wings. 
For A-type stars, EWaw agrees well with EW12 and EW24 where EW is less than 4\AA\ and where EW is larger than 7\AA, respectively.  
EW48 and EW70  show linear correlation with EWaw, but systematically higher, indicating that EWaw could cover most of the width range, but with slightly underestimate. 
For the case of F-type stars, EW12 agrees well width EWaw where EW is less than 4\AA\ and begin to be underestimated as EW grows. 
Since    H$\alpha$ is weaker but the metallic line in the wings are stronger than A type stars,  EW24, EW48 and EW70 obviously could not properly measure the real EW and the EW grows larger when more line wings are included.
From the figures, we can see that neither of the four fixed width could properly measure the EW considering the various width of the H$\alpha$ line. 
By using adaptive width we could measure the line EW in different  width with higher accuracy and  smaller uncertainty.  
 
\begin{figure}[bthp]
\centering
\includegraphics[width=0.7\textwidth]{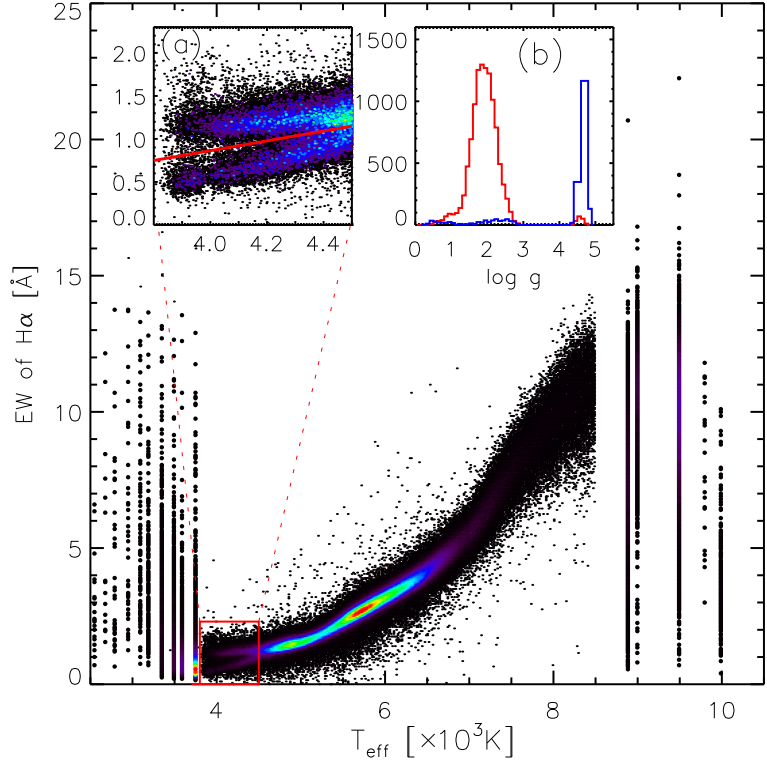}
\caption{ $T_{\mathrm{eff}}$ vs  EW of H$\alpha$ for stars in DR5.
The color indicates the number density of the stars.
Stars with $T_{\mathrm{eff}}$ between 3800 K and 4500 K are amplified in sub-panel (a), the corresponding  distribution of $\mathrm{log}\textit{ g}$ of the stars separated by the red line are shown in sub-panel(b).  
} \label{ewteff}
\end{figure}

 A uniformly defined EW is more convenient to do comparison between different elements and stellar types.
As shown in Figure \ref{ewteff}, there is a clear correlation between   EW of H$\alpha$ and effective temperature of all LAMOST single epoch spectra in DR5PC.
When $T_{\mathrm{eff}}$ is lower than 4300  K,  giants and dwarfs could be clearly separated by their EWs.

\subsection{Embedded Balmer  emission lines in early type star}\label{secanmlines}
For stars hotter than F, Balmer lines are more important in RV measurements than in cooler stars, because they are the dominating lines and the metal lines are weak. 
The pure emission lines are masked out during measuring the RV of a star.
Abnormal Balmer lines, such as emission line embedded in an absorption line or double absorption lines of double-line spectroscopic binaries (SB2), can lead to errors in RV determination.  
The cases of stellar absorption line mixed with emission line, which originates mainly in (1) stellar envelopes or outer stellar atmospheres, (2) stellar activities and (3) binary interaction, are more popular in abnormal Balmer lines than SB2s.
In the case of SB2, only when the brightness of the two components is similar and the RV difference is greater than the resolution, can it be observed, which requires that the mass ratio of the two components is close to 1 and they are in special phases in their orbit.
These conditions make the probability to observe SB2 very small.

For the cooler stars, the Balmer lines are weak and the line cores are  usually narrower than 100/kms (\cite{moss1981}),  and the line wings are polluted by metal lines.  
They look narrow and appear  in either pure emission or absorption in the low resolution mode, thus it is hard to separate the emission from the absorption line.  
Also the abnormal Balmer lines will not lead to errors in the RV measurement, since there are enough  metal lines even when the hydrogen lines are masked.
Due to these reasons, we do not try to identify the second component of Balmer lines for stars later than F-type.

We use the following procedure to identify the abnormal H$\alpha$, H$\beta$, H$\gamma$ and H$\delta$ for the stars earlier than or equal to the F-type.
First of all, an initial line range is defined depending on the stellar type.
Because A-type stars have the  strongest Balmer lines among all types, a wider range is applied to A star than others.
The range for O, B, A, and F types is set to be 60, 60, 100, 60\AA, respectively.
Given a spectrum, the local pseudo-continuum of the Balmer line is derived by fitting a fourth-order polynomial  to the segments about 200\AA\ beyond the initial range. 
The local pseudo-continuum is then divided  to obtain the normalized spectrum $f$.

\begin{figure}[bthp]
\centering
\includegraphics[width=0.5\textwidth]{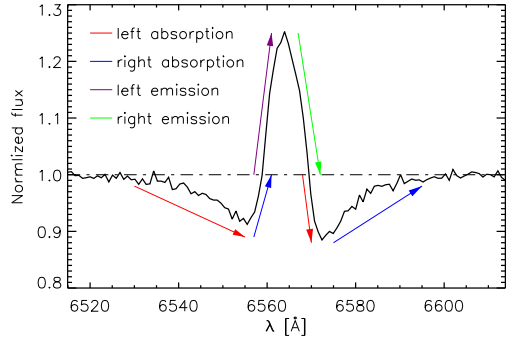}
\caption{A sample of wing detection. Red, blue, purple, green lines are the left part of an absorption, the right part of an absorption, the left part of an emission, the right part of an emission, respectively, as denoted in the figure.} \label{lwing}
\end{figure}

As indicated in Figure \ref{lwing},  a spectral line could be detected by searching the  segments of ascending and descending pairs.
Both emission and absorption contain a pair of an ascending segment and a descending segment.
For spectrum $f$, we define $\Delta{f}$  as follows,

\begin{equation}\label{deltaf}
    \Delta_i{f}= \left\{ 
    \begin{array}{rl}
    1 & ,\ \mathrm{if}\ f_{i+1}-e_{i+1}\geq f_{i}\ \mathrm{or}\ f_{i+1}\geq f_{i}\geq f_{i-1}\\
    -1 & ,\ \mathrm{if}\ f_{i+1}+e_{i+1}\leq f_{i}\ \mathrm{or}\ f_{i+1}\leq f_{i}\leq f_{i-1}\\
    0 & ,\ \mathrm{else}
    \end{array} \right.,
\end{equation}
where $e$ is the error of corresponding $f$. 
A successful detection of an(a) ascending(descending) segment requires at least 5 consecutive pixels (2.5\AA) of  $\Delta{f}=1(-1)$. 
If segments length is less than 5 pixels, $f$ is smoothed with a median filter of 3-5 pixels and $\Delta{f}$ is recalculated to see if the 5-pixels-length condition could be met. 
If  more than one pairs of segments are detected within the wavelength range, the line is very likely to be abnormal.  
Segments are combined into pairs according to  Table \ref{linewing}. 
Each pair is first fitted  by a S\'ersic profile $\hat{f}$, then subtracted from  $f$.
The residual  $r=f-\hat{f}$ is then fitted by a new S\'ersic profile, $\hat{r}$, so the fitted profile is $f-\hat{f}-\hat{r}$. 
The goodness of the fit is evaluated by the reduced $\chi^2$. 
Among all combination of pairs, the one with the smallest   reduced $\chi^2$ is considered as the best fit. 

\begin{table}[bthp]
\begin{center}
\centering \caption{The wing detection of a spectral line}\label{linewing}
\begin{tabular}{c|cc|cc}
\hline\hline
type & emission & & absorption & \\
\hline
side & left & right & left & right \\
\hline
$f$ & $>1$ & $>1$ & $<1$ & $<1$ \\
$\Delta{f}$ & $>0$ & $<0$ & $<0$ & $>0$ \\
\hline
\end{tabular}
\end{center}
\end{table}

\begin{figure}[bthp]
\centering
\includegraphics[width=0.5\textwidth]{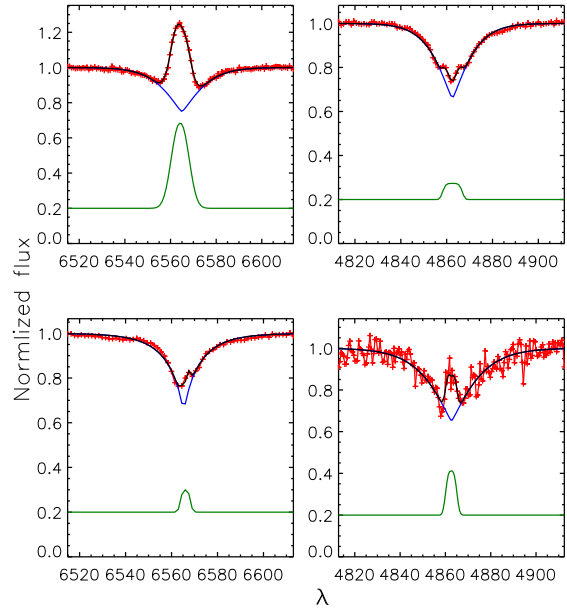}
\caption{Four samples of line fit. In each panel, the red, blue, green and black lines imply the normalized spectra, the fitted absorption line, the emission component and the fitted combination of the two components, respectively.} \label{lfsam}
\end{figure}

It turns out that the combination of  an emission and an absorption line always give a better fitting than double absorptions. 
This agrees with that in the low resolution mode most binaries are single-lined spectra. 
But caution must be taken that  some real double-lined binaries might be misidentified as an embedded emission and an absorption. 
This is because  two absorption solution usually includes a pair of unsymmetrical segments,  leading to larger residual than using combination of symmetric emission and absorption profile. 
So for these relative rare cases, current fitting might be wrong. 
We leave the disentangling of double-lined binary for the future work. 
 The outer wing of the fitted profile and the central part of the actual spectrum are used to calculate the EW, similar to Section \ref{secnmlines}. 
The RV of the emission and the absorption  are  measured respectively. 
Figure \ref{lfsam} shows several examples of  the absorption and  emission combo, in which the first sample is the fit of Figure \ref{lwing}.

Balmer emission line can have complex structure, such as Smak  profile of a disk (\cite{Smak69})
 or P Cygni profile of strong stellar wind, although most of these profiles will not
 show too much difference with a single emission line added on an absorption in the low resolution mode.
 We did not try to  decompose  the different  components of these complex Balmer lines.
The primary purpose of the detection is to measure EW correctly and remove the abnormal Balmer bands in the following calculation of RV.

\section {Radial velocities}

RV is the basic information to describe the motion of a star.
In this section we describe the RV measurement for LAMOST single epoch data, 
using the classical cross-correlation technique, 
i.e., by shifting and comparing the best matched template to the observed spectrum. 

\subsection {Band selection}                                                          
The wavelength bands  to calculate the cross-correlation function (CCF) for  stars hotter than G stars are 4000-5300\AA\ and 6350-7000\AA,  for those cooler than K are 4550-5300\AA\ and 6350-9000\AA. 
To remove the pixels with low SNR, only the segments that have more than 40 consecutive pixels with SNR greater than 5 are retained.
The obvious DiB bands (4431\AA, 5782\AA, 5798\AA, 6285\AA, 6616\AA), telluric bands (A band at 7560$\sim$7720\AA, B band at 6850$\sim$6960\AA, as well as 7150$\sim$7350\AA and 8105$\sim$8240\AA), strong sky emission lines, nebular lines and the detected emission lines (described in Section \ref{secnmlines} and \ref{secanmlines}) are masked in the calculation.
Note that caution should be taken when using data earlier than 2012 Jan 14. 
Due to a problem  with Ne/Ar lamp, only the wavelength  calibration in the blue part is reliable. 
For these data, only 4000-5300\AA\  band is used to measure RVs for all type of stars. 
The RVs might have large uncertainty for late type stars or spectra with low SNR in that wavelength range. 

\subsection {Templates\label{sec:temp}}
\begin{figure}[bthp]
\centering
\includegraphics[width=0.5\textwidth]{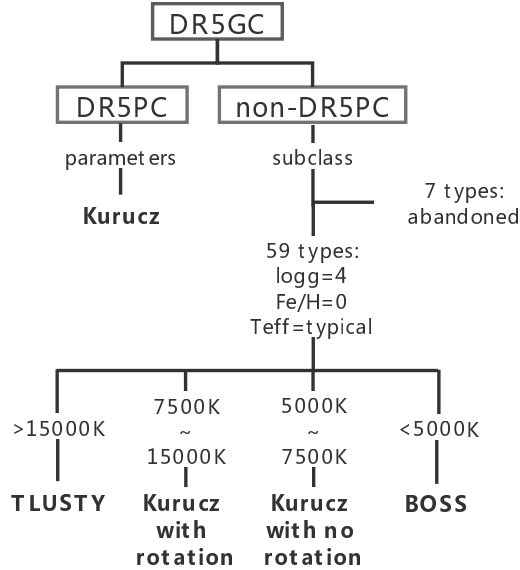}
\caption{Templates used for RV measurements according to their $T_{\mathrm{eff}}$ and stellar type.}
\label{rvtemplate}
\end{figure}

 The overall  template selection scheme is summarized in  Figure \ref{rvtemplate}.
 For each star, an appropriate template is chosen  according to the spectral classification   of 1D pipeline. For the stars in DR5PC, since the basic stellar atmosphere parameters are provided, 
the templates are constructed from  the corresponding Kurucz model with resolution degraded to R$\sim$1800.
For  other stars in DR5GC, the stellar parameters are not available.  
Since  the CCF method is not sensitive to  $\mathrm{log}\textit{ g}$ and $[\mathrm{Fe/H}]$ in the low resolution mode,
 their $\mathrm{log}\textit{ g}$ and $[\mathrm{Fe/H}]$ are fixed to  4 and 0, respectively, and $T_{\mathrm{eff}}$ is determined by their stellar sub-class (\cite{bookssc}), as  in Table \ref{typicalteff}.
For  seven special  types listed in DR5GC, including Carbon, CarbonWD, CV, DoubleStar, WD, WDMagnetic and Non, due to lack of corresponding template, the  RVs are not released,  only the  single epoch spectra are available.
For the remaining stars, various stellar atmosphere template are used according to their typical $T_{\mathrm{eff}}$.
For stars with typical $T_{\mathrm{eff}}$ higher than 15000 K, the templates are generated by  TLUSTY model of hot stars (\cite{lanz2007}) with  a set of $v\mathrm{sin}i$ from 100  km s$^{-1}$ to  200  km s$^{-1}$ in a step of 20 km s$^{-1}$.
The same $v\mathrm{sin}i$  grids are constructed  for templates of stars with $T_{\mathrm{eff}}$ in the range of 7500K to 15000K, but with the Kurucz model.
For  $T_{\mathrm{eff}}$ between 5000K and 7500K, we use Kurucz templates with no rotation. 
For stars cooler than 5000K, the dwarf model from BOSS (\cite{kesseli2017}) is adopted.

\begin{table}[bthp]
\begin{center}
\centering \caption{The typical $T_{\mathrm{eff}}$ for star sub-classifications}\label{typicalteff}
\begin{tabular}{ll|ll|ll|ll}
\hline\hline
subclass & $T_{\mathrm{eff}}$ (K) & subclass & $T_{\mathrm{eff}}$ (K) & subclass & $T_{\mathrm{eff}}$ (K)  & subclass & $T_{\mathrm{eff}}$ (K) \\
\hline
A0         & 9800   &  A7V        & 7800   &  G0         & 5900  & K5         & 4400    \\
A0III      & 10000  &  A8III      & 7580   &  G1         & 5800  & K7         & 4130    \\
A1IV       & 9500   &  A9         & 7450   &  G2         & 5750  & M0         & 3750    \\
A1V        & 9500   &  A9V        & 7380   &  G3         & 5700  & M1         & 3600    \\
A2IV       & 9000   &  B6         & 14000  &  G4         & 5640  & M2         & 3500    \\
A2V        & 8900   &  B9         & 10700  &  G5         & 5580  & M3         & 3350    \\
A3IV       & 8500   &  F0         & 7250   &  G6         & 5510  & M4         & 3200    \\
A3V        & 8500   &  F2         & 7000   &  G7         & 5470  & M5         & 3100    \\
A5         & 8150   &  F3         & 6750   &  G8         & 5430  & M6         & 2950    \\
A5V        & 8150   &  F4         & 6650   &  G9         & 5350  & M7         & 2800    \\
A6IV       & 7900   &  F5         & 6550   &  K0         & 5280  & M8         & 2680    \\
A6V        & 8000   &  F6         & 6400   &  K1         & 5110  & M9         & 2550    \\
A7         & 7800   &  F7         & 6250   &  K2         & 4940  & O          & 40000   \\
A7III      & 7750   &  F8         & 6170   &  K3         & 4700  & OB         & 30000   \\
A7IV       & 7750   &  F9         & 6010   &  K4         & 4550  &            &         \\
\hline
\end{tabular}
\end{center}
\end{table}

\subsection {RV Determination} 
\begin{figure}[bthp]
\centering
\includegraphics[width=0.5\textwidth]{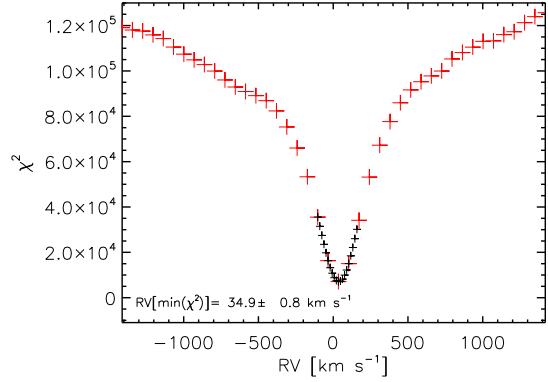}
\caption{The RV grids to determine the minimum of the $\chi^2$.
 The large and small symbols denote the RV grids in the first and the second round, respectively.} \label{rvsolve}
\end{figure}

RV is calculated by minimization of $\chi^2$  between the normalized spectra  and the RV-shifted normalized template.
To reduce the computing time, the calculation is performed in two rounds.
In the first round, the RV grid added to the template is -1500 km s$^{-1}$$\sim$1500 km s$^{-1}$ with a step of $\Delta{\log\lambda}=0.0001$ ($\Delta{v}=69$ km s$^{-1}$).
Then $v_b$, the RV with the minimum $\chi^2$,  is selected as the peak of the first round.
In the second round, in the range of $v_b-138$ km s$^{-1}$$\sim{}v_b+138$ km s$^{-1}$, we use a grid of $\Delta{\log\lambda}=0.00002$ ($\Delta{v}=13.8$ km s$^{-1}$), as shown in Figure \ref{rvsolve}.
The bottom of the $\chi^2$ curve is fitted by a Gaussian function to find the minimum  and the fitting error.
The minimum is regarded as the final RV. 
The error of the final RV is the square root of the sum of the squares of two factors. 
One is the fitting error of the profile center of the Gaussian function  and the other is the formal error where $\Delta\chi^2=1$. 

\subsection{RV accuracy}
\begin{figure}[bthp]
\centering
\includegraphics[width=0.5\textwidth]{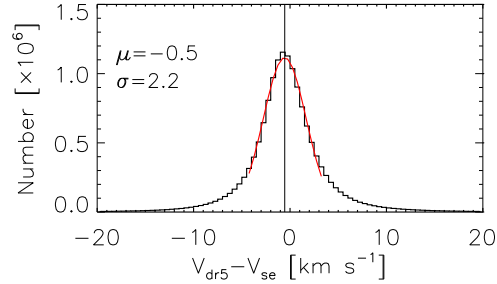}
\caption{The comparison of RVs derived from this work and  those from the parameter catalog of LAMOST DR5.
The red Gaussian profile is the fit of the core of the histogram, 
which has an average value of -0.5  km s$^{-1}$ and a standard deviation of 2.2  km s$^{-1}$.}
\label{rvvsDR5}
\end{figure}

We first compare our single epoch RVs with those released in DR5PC, for which the
co-added spectral SNR of the $g$ band is required to be greater than 20. The results is  shown in Figure \ref{rvvsDR5}.
The RV of this work is 0.5  km s$^{-1}$ larger than the RV in DR5PC, with a standard deviation of 2.2  km s$^{-1}$. 
Considering the resolution unit is about 170 km s$^{-1}$, there is no system difference between the DR5PC RV and the RV in this work.

The RV differences between different exposures of the same star are used to estimate the RV measurement error.
The RV uncertainty depends on the number and width of the absorption lines in the spectrum, so the most important factor affecting the RV accuracy is the spectrum type,
although $\mathrm{log}\textit{ g}$ and $[\mathrm{Fe/H}]$ also play a smaller role.
Due to the relatively small size of  the hot star samples, only stars of A, F, G, K and M types are included in the statistics.
For different exposures on the same day, the typical time span between different exposures is less than 2 hours, as shown in Figure \ref{obssta}. 
The same object is usually observed in the same fiber  and the instruments are stable within the short period. 
While for exposures in different days, the situation is more complex. 
Many reasons can result in the difference in RV measurement. 
For example, the temperature difference could cause the shift of the spectrograph and a slow decay of the image quality. 
The profile difference between different fibers in different date  leads to a wavelength calibration  error of a few  km s$^{-1}$.  
To better represent the actual RV error, we distinguish the RV differences on the same day from those in different days.

\begin{figure}[bthp]
\centering
\includegraphics[width=0.5\textwidth]{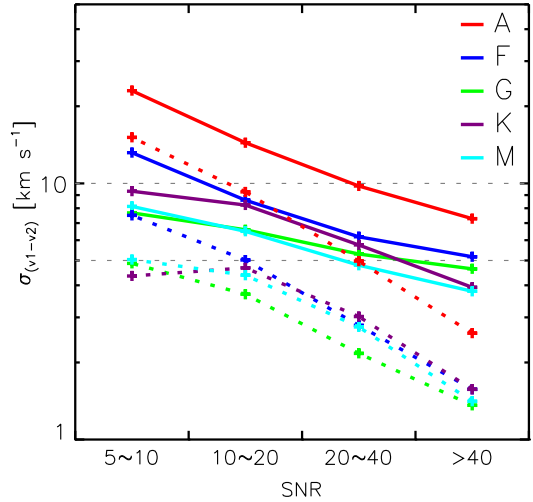}
\caption{The standard deviation of RV difference between two exposures depending on stellar types and SNR. 
Dotted lines and solid lines are for the RV difference within one day and between days, respectively.
v1 and v2 in the title of y-axis denote RV obtained from two exposures of the same star.
Colors are used to distinguish different stellar types as marked in the figure.}
\label{drvexpall}
\end{figure}

We make statistics of the RV difference depending on the spectral types under different SNR ranges.
For a certain range, only the two spectra whose SNR are both in this range can participate in the statistics.
No systematic difference is found between exposures for all type of stars, indicating that the wavelength calibration is stable over the first stage of the LAMOST survey. 
The standard deviations are shown in Figure \ref{drvexpall} and Table \ref{tblrverr}.
As can be seen in Figure \ref{drvexpall}, hotter stars have larger scatter since there are fewer   absorption lines and the lines are  broader as the temperature increases.
It also could be noticed that the RV accuracy increases with the SNR of the spectrum.  
For the spectra with SNR greater than 10, the deviations from different days are less than 10  km s$^{-1}$ for all types but A stars, for which the scatter is  14.41  km s$^{-1}$ when the SNR is between 10 and 20. For G, K and M stars, the RV accuracy is higher than 10  km s$^{-1}$ when SNR is higher than 5.
Figure \ref{drvexpall} and Table \ref{tblrverr} could be used as a standard to select RV variation stars.
Because SNR of both the two spectra are in the same range, their RV uncertainties are in the same level, so that the RV difference is approximately $\sqrt{2}$ times the RV uncertainty.

\begin{table}[bthp]
\begin{center}
\centering \caption{The standard deviation of $\Delta$RV (in  km s$^{-1}$) between two exposures. 
Exposures in the same day are in brackets, exposures between days are outside brackets.}\label{tblrverr}
\begin{tabular}{c|ccccc}
\hline\hline
SNR & A & F & G & K & M \\
\hline
$5 \sim 10 $ & 23.01(15.12) & 13.18(7.50) & 7.67(4.87) & 9.32(4.35) & 8.13(5.04) \\
$10 \sim 20$ & 14.41(9.26) & 8.60(5.02) & 6.59(3.69) & 8.23(4.68) & 6.50(4.38) \\
$20 \sim 40$ & 9.77(4.97) & 6.19(2.79) & 5.30(2.17) & 5.75(3.02) & 4.79(2.75) \\
$ > 40  $ & 7.28(2.60) & 5.17(1.57) & 4.64(1.36) & 3.93(1.57) & 3.79(1.41) \\
\hline
\end{tabular}
\end{center}
\end{table}

\subsection{Caveats}

There are a few issues needed to be aware of when using the data.
First, the pollution from the bright adjacent fibers could affect the RV measurement.
The   spectra  polluted by their bright neighbor usually have obvious irregular features, such as abnormal continuum, large residual of the telluric bands, or wrong spectral lines. 
One can use the ratio of the flux to the stronger neighbor (included in RV data) to see if the fiber is possibly affected. 
Second, early type stars which have strong Balmer emission lines may have bad RV measurements since hot stars are lack of  other  absorption lines.
Once the emission component is detected in H$\alpha$,   other Balmer absorption lines could  also be polluted  even if no emission component is detectable, leaving the RV measurement to be biased.
Third, the RV may be wrong due to the strong sky background if the exposures are taken in the condition that is not good for astronomical observation, e.g. in nearly full moon nights,  too close to the Moon, or beyond astronomical twilight.
Information about the observation, including the position of the Sun, the Moon as well as the Moon phase, can be browsed online.

\section {Data products}\label{sec:product}

The current data release includes all FITS file of  single epoch spectra,  information of repeat observation of individual targets, measured RV and EW catalog for stellar objects in LAMOST DR5. 
All the data could be queried, browsed  and downloaded online\footnotemark.
\footnotetext{http://dr5.lamost.org/sedr5/}

All the spectra in FITS format, are distributed in the sub-directories of \textit{obsdate} and \textit{planid}, 
and the file name is consisted of \textit{obsdate}+\textit{planid}+\textit{spid}+\textit{fiberid}, 
where  \textit{obsdate} denotes the observation date, 
\textit{planid} denotes the name of the plate,  
\textit{spid} denotes the ID of the spectrograph (1-16) and  
\textit{fiberid} denotes the ID of fiber (1-250) in the same spectrograph.
The spectrum of individual exposures in the same observation are wrapped in the same file.
For example, the file '20111024/F5902/20111024\_F5902\_01\_001.fit' contains spectra of all the sub-exposure and information of the first fiber of the first spectrograph in the \textit{planid} of F5902 taken on 2011 Oct 24. 
The header of each file contains the information of the observation, the instruments and the data reduction, etc.

Each file contains three extensions of ASCII table.
The first extension contains the spectra, described as follows: 
\begin{itemize}
\item \textbf{flux}: flux in counts per pixel, an array with the size of 4136*$n$, where $n$ is the number of individual frame. Usually $n$ is twice the number of exposures including the red and blue branch of the spectrum, but there are rare cases that one part of the spectrum fails, then users should refer to \textbf{color} and \textbf{mjm} to know which part is lost. 
\item \textbf{invvar}: invert variance, an array with the size of 4136*$n$.
\item \textbf{loglam}: logarithm of wavelength in \AA, an array with the size of 4136*$n$. The wavelengths are in vacuum scale and heliocentric frame.
\item \textbf{pixelmask}: pixel mask, an array with the size of 4136*$n$.
\item \textbf{skyflux}: subtracted sky flux in counts per pixel, an array with the size of 4136*$n$.
\item \textbf{fluxcorr}: relative flux calibration curve, an array with the size of 4136*$n$.
\item \textbf{mjm}: the time at the beginning of the exposure, an array with the size of $n$. The 8 digits \textit{mjm} means the number of minutes elapsed from 0:00 Nov 17th 1858 Beijing time to the beginning of the exposure.
\item \textbf{color}: branch, {\it b} or {\it r}, an array with the size of $n$.
\item \textbf{exptime}: exposure duration time in second, an array with the size of $n$.
\item \textbf{fibermask}: fiber mask, an array with the size of $n$.
\end{itemize}
The second and the third extensions contain information of arc lines in the blue and red branches, respectively. 
The arc line information is useful to determine  the dispersion profile of the instrument. The structure of each extension is:
\begin{itemize}  
\item \textbf{line}: the wavelength of used arc lamp lines, in \AA 
\item \textbf{xarc}: the pixel position of the line centers in the raw image, in pixel 
\item \textbf{width}: the FWHM of the lines, in \AA 
\item \textbf{inten}: the intensity of the lines, in counts per pixel
\item \textbf{werror}:  the wavelength error, in \AA 
\end{itemize}

DR5 contains 8 million stellar spectra, and some of them are from the same target.. 
Since the typical seeing and the size of LAMOST fibers are both about 3 arc-seconds, targets within 3 arc-seconds are considered as the same one. 
As a result, all stellar spectra come from 6.5 million targets.
A unique ID, \textit{gid}, is used to mark the same target  in different observations. 
For each stellar target, we counted its exposure times, observation times, RV difference and other information. 

The information of 6.5 million stellar targets,
RV and EW for 7.7 million stellar spectra (more than 20 million RV records and more than 350 million EW records) are stored in three tables of the database, which could be accessed through the web page. 
These tables are:
\begin{enumerate}
\item \textbf{reobsstat}: information of repeat observation of individual targets, such as number of repeat exposures, maximum RV difference between difference exposures, etc.
\item \textbf{sexprv}: RV measurements of  individual single epoch spectrum.
\item \textbf{lineindices}: EW of spectral lines in individual single epoch spectrum.
\end{enumerate}

The detailed description of individual items in the tables as query examples can be found  online and will not be repeated here.
Targets in the three tables could be cross identified by  \textit{gid}.  
For the same targets, information of different single epoch spectra such as RV or EW could be found under the same \textit{gid}.
One can use  \textit{gid} to obtain  \textit{obsdate}, \textit{planid}, \textit{spid} and \textit{fiberid} to access the FITS file as described previously.

\section {Summary}
We release the single epoch data of LAMOST DR5, including  25.05 million  low resolution spectra of 6.56 million targets, covering 17000 deg$^2$  of the first stage of the LAMOST survey. 
The data were processed by the LAMOST 2D pipeline, where they  were sky subtracted, cosmic ray rejected and wavelength calibrated, but not flux corrected. 
For each stellar spectrum, we measured the EWs of 60 lines of 11 elements with a new method combining the actual line core and the analytic function of the fitted line wing to further depress the noise.  
For early type stars, the Balmer lines with   embedded emission feature were fitted with two components of absorption and emission, which could help to understand the origin of both components. 
RV of each single spectrum was derived by cross correlating  the spectrum with its best matched template.  
RV accuracy is a function of spectral type and SNR, could also depend on the time lag between observations due to the unknown instability of the instruments, for which the LAMOST team will continue to refine the algorithm and spectra quality.
All the spectral FITS files and results of 20.94 millions RVs and 376.44 millions EWs are available online.  
The RVs and EWs are stored in the database and could be browsed and queried using the interface. 

The single epoch data extend the LAMOST time domain spectra to shorter time scale of tens minutes , so that the RV of close binaries with period of hours to days could be discernible in the single epoch data.  
Close binary candidates based on the current database will be available at Yuan et al., in preparation. 
Combining the spectral information of the primary mass and RVs, the sample will give better constraint on the mass of the secondary,  and offer a larger sample with better mass estimation to understand the evolution of close binaries.

\textbf{Acknowledgments}
ZHT acknowledges the support of the National Key R\&D Program of China (2019YFA0405000), the National Natural Science Foundation of China (NSFC) (grant no. 12090040 and 12090041).
BZR acknowledges the support of NSFC (grant no. 11973054).
YHL acknowledges the support of the Youth Innovation Promotion Association of Chinese Academy of Sciences (id. 2020060).
This work has been supported by Cultivation Project for LAMOST Scientific Payoff and Research Achievement of CAMS-CAS.
Guo Shou Jing Telescope (the Large sky Area Multi-Object fiber Spectroscopic Telescope, LAMOST) is a National Major Scientific Project built by the Chinese Academy of Sciences. Funding for the project has been provided by the National Development and Reform Commission. LAMOST is operated and managed by the National Astronomical Observatories, Chinese Academy of Sciences.


\end{document}